\documentclass[journal=jacsat,manuscript=article]{achemso}

\usepackage[version=3]{mhchem} 

\author{Esita Pandey}

\author{Shaktiranjan Mohanty}

\author{Abhisek Mishra}

\author{Bhuvneshwari Sharma}

\author{Subhankar Bedanta}
\affiliation{Laboratory for Nanomagnetism and Magnetic Materials (LNMM), School of Physical Sciences, National Institute of Science Education and Research (NISER), An OCC of Homi Bhabha National Institute (HBNI), Jatni 752050, Odisha, India}
\email{sbedanta@niser.ac.in}

\title[An \textsf{achemso} demo]
  {Structural deformation and irreversible magnetic properties of flexible Co/Pt and Co/Pd thin films}


\begin{document}
\begin{abstract}
The successful commercialization of flexible spintronic devices requires a complete understanding of the impact of external strain on the structural, electronic, and magnetic properties of a system.  The impact of bending-induced strain on flexible films is studied quite well. However, little is known about the effect of other modes of flexibility, e.g., wrinkling, twisting, peeling, and stretching on the functional properties of flexible films. In this context, perpendicular magnetic anisotropic Co/Pt and Co/Pd thin films are prepared on flexible Kapton substrates, and the impact of the peeling mode is studied in detail. The peeling method generates numerous cracks, and buckling in the thin film, along with localized blister formation imaged by scanning electron microscopy. Further, the resistivity measurement confirms a significant enhancement in sample resistance owing to the severe damage of the films. The structural discontinuities strongly affect the magnetization reversal phenomena as measured by the magneto-optic Kerr effect (MOKE)-based microscopy. The bubble domains got converted to elongated-shaped domains due to several hindrances to the wall motion after strain application. Further, the relaxation measurements reveal that the thermal energy is insufficient to switch the magnetization at a few areas due to their high pinning potential associated with the damages. In contrast to bending-induced strain, here, all the modifications in the functional properties are found to be irreversible in nature.
\end{abstract}

\section{Introduction}
Comprehending the effect of external strain on the structural, electrical, and magnetic properties of flexible films is crucial for the successful commercialization of flexible spintronic devices. Different methods to generate strain on a flexible film are, bending, stretching, twisting, peeling, wrinkling, etc \cite{chen2021mechanically,bao2016flexible,duan2015highly,huang2018conformal,yang2018effect,hong2023real,cheng2022strain}. Among them the bending and stretching modes generate mainly uniaxial strain, whereas the other methods may generate a multi-dimensional strain on the system \cite{yang2018effect,hong2023real}. For wearable device applications, the flexible films should withstand all the above-mentioned modes of flexibility \cite{yang2018ionic, liu2018mechanical, ha2018wearable}. The effect of bending-induced strain is studied quite well for both the in-plane (IP) and perpendicular anisotropic thin films \cite{sheng2018flexible, shen2023flexible, pandey2020strain, zhao2018low}. Within the elastic limit of bending strain the structural and magnetic properties are usually found to be reversible in nature. However, it is important to know the critical strain beyond which severe deformation may take place. The occurrence of crack formation and buckle delamination may strongly affect the functional properties of a film. This may restrict the industrialization of flexible devices and necessitates a thorough understanding. However, these topics are rarely studied in detail. There are only a few reports that focused on studying the effect of stretching, twisting, and wrinkling modes on the structural and/or magnetic properties of a film \cite{an2020highly,li2016stretchable,liu2021stretching,faurie2017fragmentation,merabtine2018origin}. The response to such strain depends on the magnetostrictive coefficient ($\lambda$) of the magnetic film. Faurie et al. studied the effect of applying large tensile strain on the Co$_{40}$Fe$_{40}$B$_{20}$ ($\lambda$$\neq0$) film deposited on the Kapton substrate \cite{faurie2017fragmentation}. Crack initiation occurred from $\sim$1.6$\%$ strain. Cracking and subsequent buckling increased the electrical resistance by several orders of magnitude. Further, Merabtine et al. studied the effect of large stretching strain on the magnetic properties of Co$_{40}$Fe$_{40}$B$_{20}$ ($\lambda$$\neq0$) and Ni$_{80}$Fe$_{20}$ ($\lambda$ $\approx 0$) thin films \cite{merabtine2018origin}. Here, crack formation started from $\sim $2$\%$ strain, whereas blister kind pattern appears from $\sim $6$\%$ strain. The stress generated during the damage is found to affect the magnetic properties, whereas the structural discontinuities are found to have no significant impact\cite{merabtine2018origin}. However, it is intuitive that the damages should act as strong pinning sites for the domain propagation events. This may strongly affect the overall magnetization reversal and relaxation mechanism of a film. As only a few studies focused on investigating the impact of different modes of flexibility, significant research attention is required to understand the effect of structural damage formation on the magnetic properties of a flexible film. In this context, we have fabricated perpendicular magnetic anisotropic (PMA) Co/Pt and Co/Pd films on Kapton substrates and studied the impact of the peeling mode of flexibility in detail. As the heavy metal (HM)/ferromagnet (FM) based PMA systems have significant potential in spintronics and sensor-based healthcare applications, a flexible counterpart of them is taken as a model system in this work. A significant damage formation (cracking, buckling, and blister pattern) due to external strain is quite evident by scanning the sample surface. Further, a resistivity measurement revealed significant enhancement in sample resistance upon application of strain. The coercivity ($H_C$) of the samples increased whereas the domain propagation was hindered strongly by the structural discontinuities that act as a potential barrier. Further, the relaxation measurements show a slower reversal owing to the restricted DW motion after peeling the film. Thus, the structural discontinuities were found to affect the local magnetic properties (e.g., magnetization reversal, domain and relaxation dynamics, etc.) significantly.

\section{Experimental details}
Pt/Co/Pt and Pd/Co/Pd thin films are prepared upon Kapton ($\sim$50$\mu$m thick) substrate having an adhesive back. This adhesion helps to generate a strain as explained in the next section. The Kapton is stuck on a Si substrate and loaded inside the deposition chamber for sample fabrication. Here Si substrate is used as a flat base for mounting the flexible Kapton. The sample structure is shown schematically in Fig. 1. Base pressure of the deposition chamber was $\sim$ $6\times10^{-8}$ mbar. All the metallic layers are deposited via the DC magnetron sputtering technique in a high vacuum multi-deposition chamber manufactured by Mantis Deposition Ltd. UK. The deposition rates for Ta, Pt, Pd, and Co layers are maintained at 0.18, 0.15, 0.20, and 0.16 \AA/s, respectively. The substrate is rotated at 10 rpm during the sample fabrication. The structural and magnetic properties are measured in both the unstrained (as-deposited) and strained (after-peeling) states of the samples. The sample surface is scanned before and after the application of strain using a scanning electron microscopy (SEM), manufactured by ZEISS. The magnetization reversal, relaxation, and domain dynamics are studied using MOKE-based microscopy in the polar mode. The resistivity of the samples is measured using a homemade four-point probe setup.

\begin{figure}[h!]
	\centering
	\includegraphics[width=0.6\linewidth]{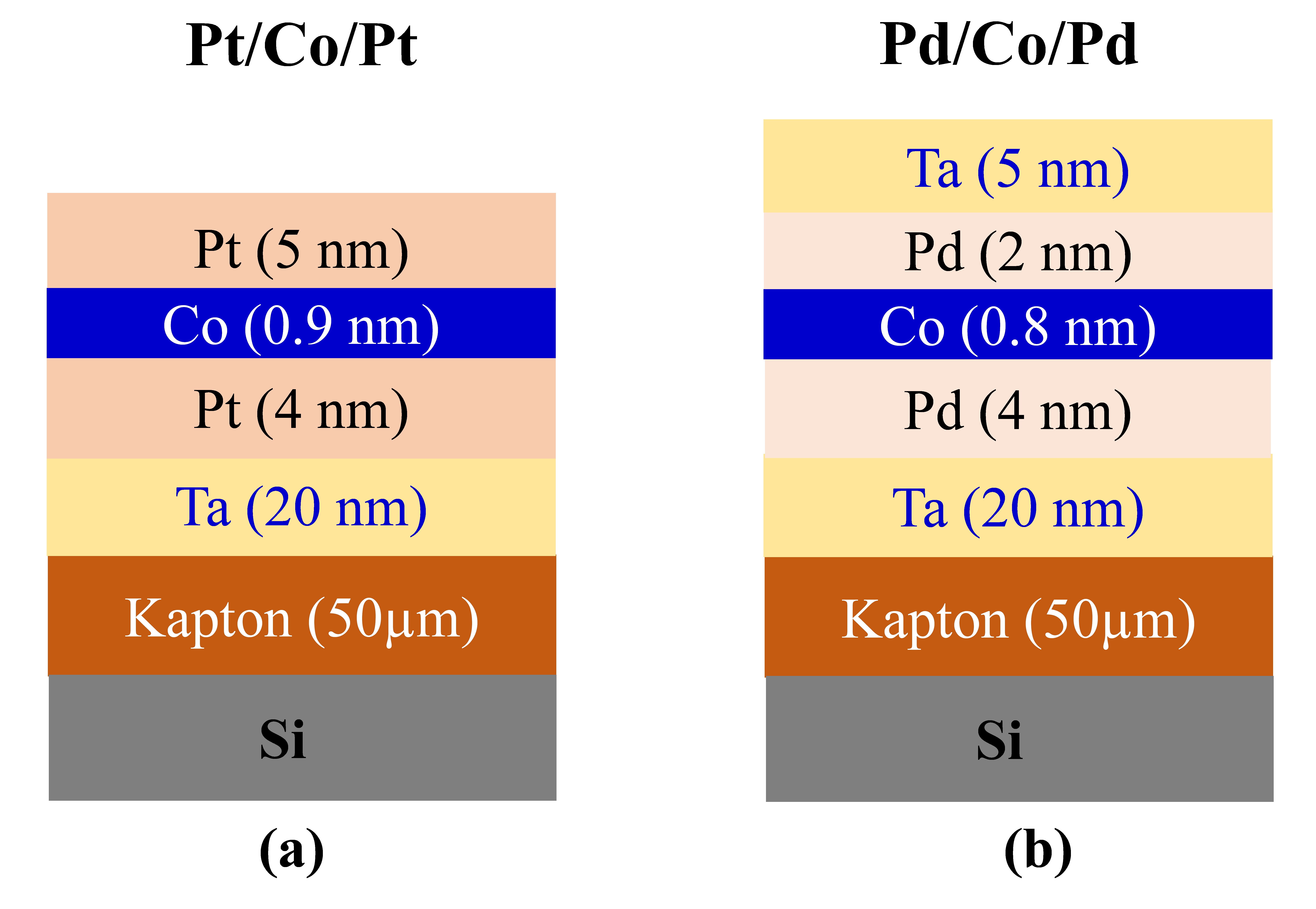}
	\caption{Schematic of the (a) Pt/Co/Pt and (b) Pd/Co/Pd sample structure deposited by the DC magnetron sputtering technique}
	\label{fig: Figure_1}
\end{figure}

\subsection{Strain generation method}
A peel-off method is used to generate strain on the deposited flexible film. In this method, the Kapton film with the sample deposited on top is peeled off from the Si substrate. Due to the adhesive back of the Kapton, the deposited film experiences a large force while detaching from the Si substrate. Along the applied force axis, it experiences a tensile stress whereas in the transverse axis, it may experience a compressive stress. The physics of such peeling-induced force was studied earlier by Lindley and Kendall in 1971 \cite{lindley1971ozone,kendall1971adhesion,kendall1975thin}. Here, by varying the strength of the applied force, the magnitude of strain can be tuned. The peel-and-stuck method is also used for fabricating flexible devices, as it is compatible with the roll-to-roll process technique \cite{huang2018conformal,lee2013peel}. A schematic representation of the strain generation method used in our work is shown in Fig. 2. For the measurements in the strained state, the samples are first peeled off from the Si substrate and then made flat to measure the properties being subjected to a high strain.  

\begin{figure}[h!]
	\centering
	\includegraphics[width=0.8\linewidth]{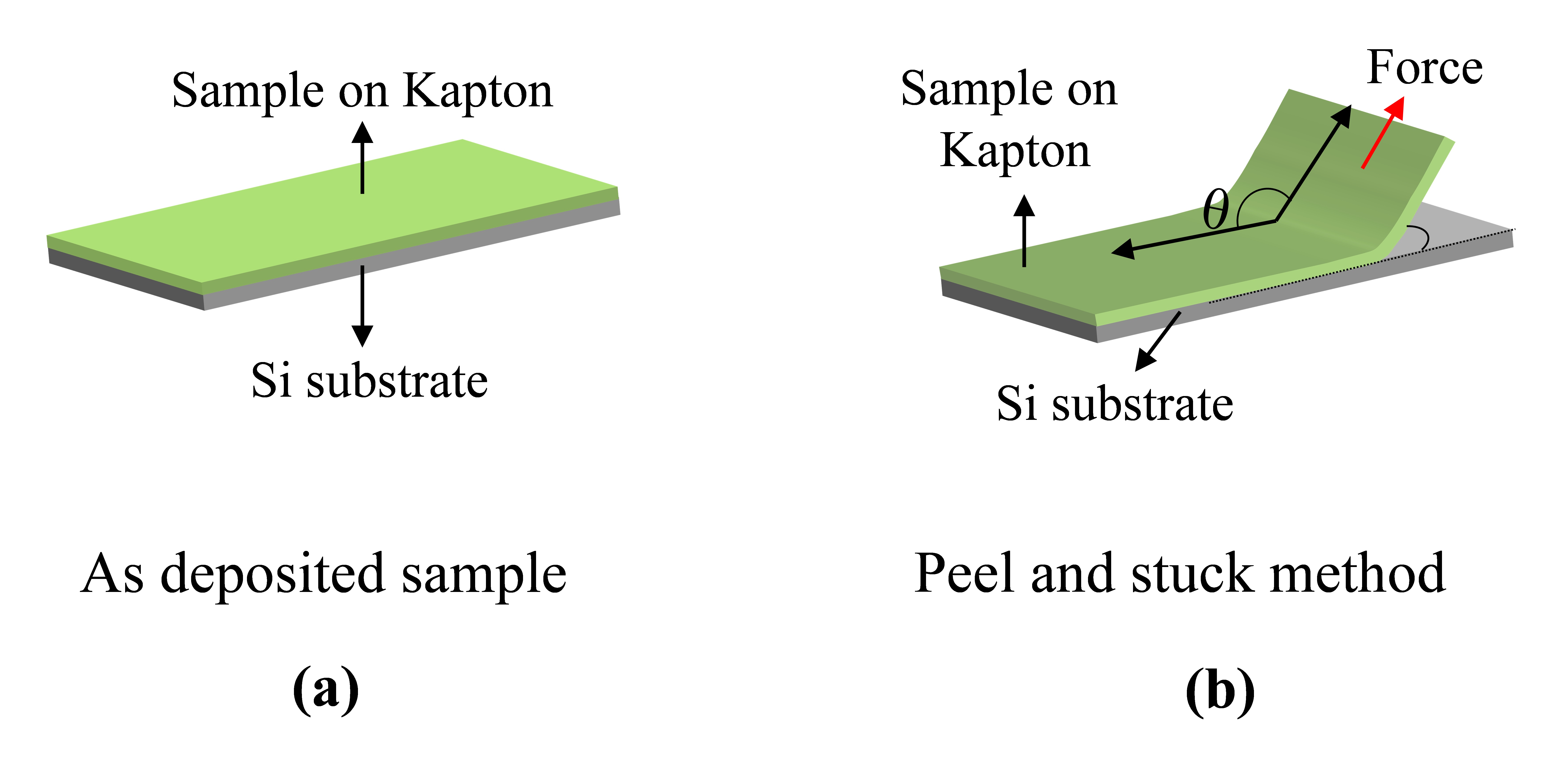}
	\caption{Schematic of the strain generation method, (a) as-deposited sample, (b) peeling off the sample from the Si base.}
	\label{fig: Figure_2}
\end{figure}

\section{Results and discussion}
Application of tensile strain initiates crack formation, whereas, a compressive strain initiates wrinkle formation and buckle delamination in the film \cite{guo2016ordered,marthelot2014self,huang2005kinetic,panin2009multi,ruoho2020thin,ni2017shape}. In the wrinkling process, both the film and substrate deform together. Thus it requires a strong adhesion and comparable stiffness between the film and substrate. However, when the film and substrate do not deform 

\begin{figure}[h!]
	\centering
	\includegraphics[width=1.0\linewidth]{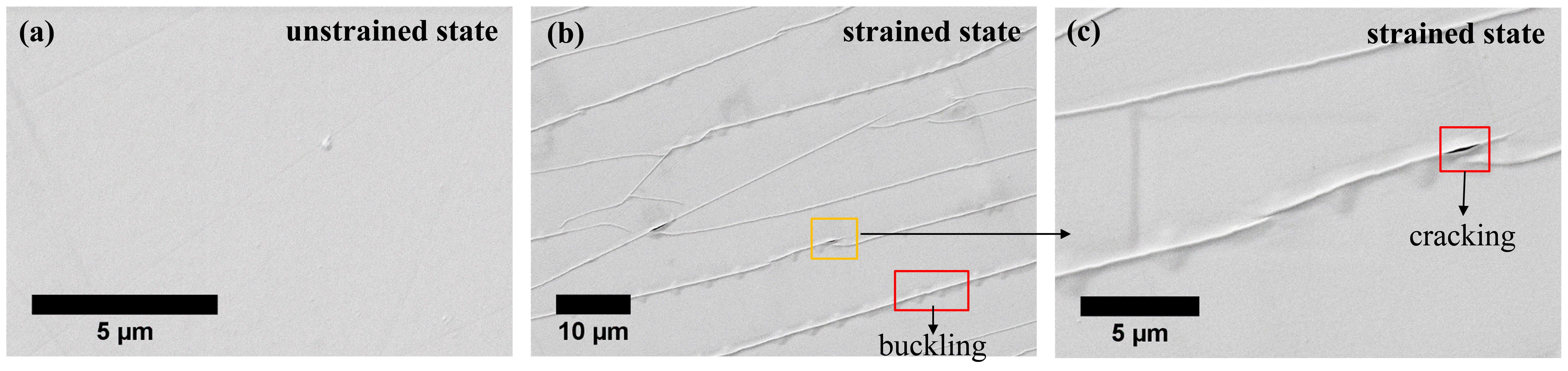}
	\caption{SEM images of the surface of flexible Pt/Co sample in the (a) unstrained, and (b-c) strained states. The red marks show the buckling and cracking of the film at several places owing to high strain. The zoom-in view of the yellow-marked region is shown in (c).}
	\label{fig: Figure_3}
\end{figure}
together, a small portion of the film detaches from the substrate to release the stress energy, known as buckling. To investigate these aspects, SEM imaging is performed for both the Pt/Co and Pd/Co samples at the unstrained and strained states of the film. Fig. 3 shows the SEM images captured for the Pt/Co sample deposited on Kapton. In the unstrained state, no sign of damage is found by scanning the sample surface. However, after peeling off several features appear on the sample surface, as shown in Fig. 3 (b). Several straight lines having small heights are observed on the film surface. This is quite similar to the buckling of thin film \cite{guo2016ordered,marthelot2014self,huang2005kinetic,panin2009multi,ruoho2020thin,ni2017shape,mei2008fracture}. Here due to the applied stress, buckling of the film is observed to minimize the elastic energy. The occurrence of buckling and wrinkling modes depends on the elastic mismatch and interfacial defects present in a system \cite{mei2008fracture}. In our case, the large mismatch between the elastic moduli of Co (209 GPa) and Kapton film (5 GPa) generates strong stress gradients at the interface of the deposited film and Kapton, resulting in the buckling of the film. At a few places, the top of the buckled film gets cracked as it experiences a high tensile strain. The red squares in Fig. 3 (b and c) represent the crack formation and buckle delamination of thin film. Fig. 3 (c) is the zoom-in view of the yellow-marked region of Fig. 3 (b). Similar to Pt/Co film, the flexible Pd/Co sample also shows no indication of damage in the unstrained state, as presented in Fig. 4 (a). After applying strain, the Pd/Co sample also exhibits distinct features on the sample surface (see Fig. 4 (b and c)). Periodic lines associated with small wavy features are found on the sample surface. These buckled portions of the film have a small height, similar to the previous case. Crack formation started from the crest of such features to release the stress energy. The red-marked region in Fig. 4 (b) shows the buckled and cracked surface of the film. A zoom-in view of such a wavy pattern is shown in Fig. 4 (c). A closer look at the red-marked region confirms the formation of a blister pattern. This pattern looks quite similar to a telephone cord blister which occurs when a film-substrate system undergoes a high compressive stress\cite{ni2017shape}. The critical stress required for such buckle formation can be calculated using the following equation \cite{freund2004thin}, 

\begin{equation}
\sigma_{b}=\frac{\pi^2}{12} \left(\frac{h}{b}\right)^2 \bar E_f
\label{2}
\end{equation}

where $h$ is the film thickness, $b$ is the half-width of the buckling, and $\bar E_f$ is the plane-strain modulus of the film. Using the value of $b$ as shown in Fig. 4 (c), the critical stress generated in our case is found to be $\sim$ 0.45 GPa. The physics of such pattern formation is quite complicated and often considered to originate from the secondary buckling of a straight-sided blister. In our case, the localized blister patterns may be associated with localized defects present at the film-substrate interface.  \cite{panin2009multi,ruoho2020thin,ni2017shape}. Thus, the crack formation, buckle delamination, and blister formation indicate a complex high-strain generation via the peeling of the flexible film. 

\begin{figure}[h!]
	\centering
	\includegraphics[width=1.0\linewidth]{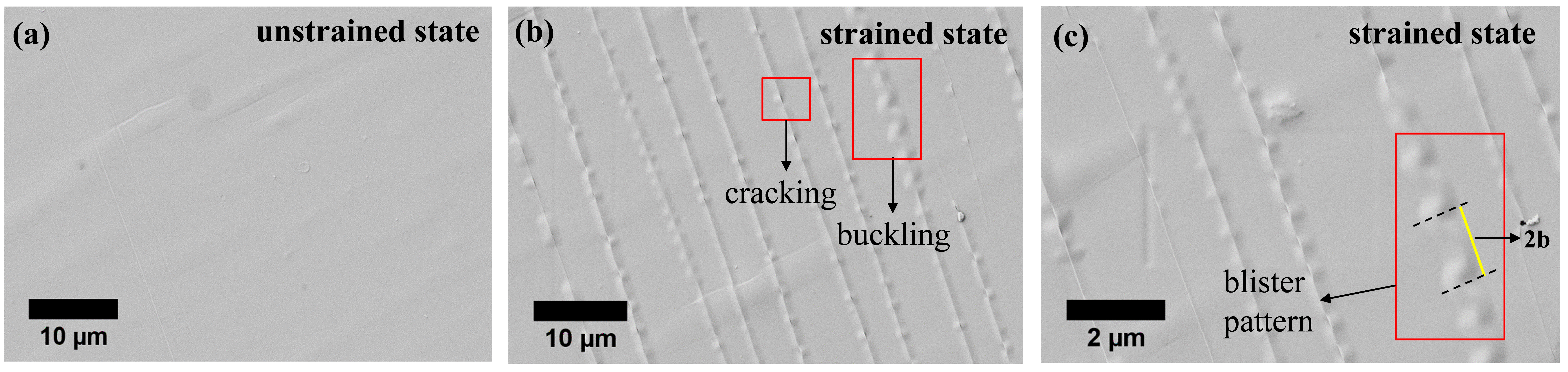}
	\caption{SEM images of the surface of flexible Pd/Co sample in the (a) unstrained, and (b-c) strained states. The red marks show the buckling and cracking of the film at several places owing to high strain. The zoom-in view of the blister pattern is shown in (c).}
	\label{fig: Figures_4}
\end{figure}

As such damage features are observed usually at high strain values, this in turn may increase the resistance of the sample. Thus, resistivity measurement is performed using a homemade four-point probe setup at room temperature (RT). Current is applied to the two outer probes whereas the voltage is measured via the two inner probes. The sheet resistance is calculated using the equation written below \cite{naftaly2021sheet},

\begin{equation}
R_S= 4.53 (\frac{\Delta V}{\Delta I})
\label{1}
\end{equation}

where, $\Delta$$V$ and $\Delta$$I$ are the output voltage and applied current to the sample, respectively. From the slope of the $V-I$ curve, the sheet resistance is calculated using equation 2. The $V-I$ curves measured in the unstrained and strained states are shown in Fig. 5. For the Pt/Co film, the sheet resistance $R_S$ was $\sim$ 45$\Omega$ in the as-deposited state, which increased to $\sim$ 800$\Omega$ after the high stress was applied. Similarly, for the Pd/Co film $R_S$ changes from $\sim$ 55 to 250$\Omega$ after the peel-off method. It indicates that the peeling mode of flexibility generates high strain on the film which increases the resistance of both the samples \cite{faurie2017fragmentation}.     

\begin{figure}[h!]
	\centering
	\includegraphics[width=0.8\linewidth]{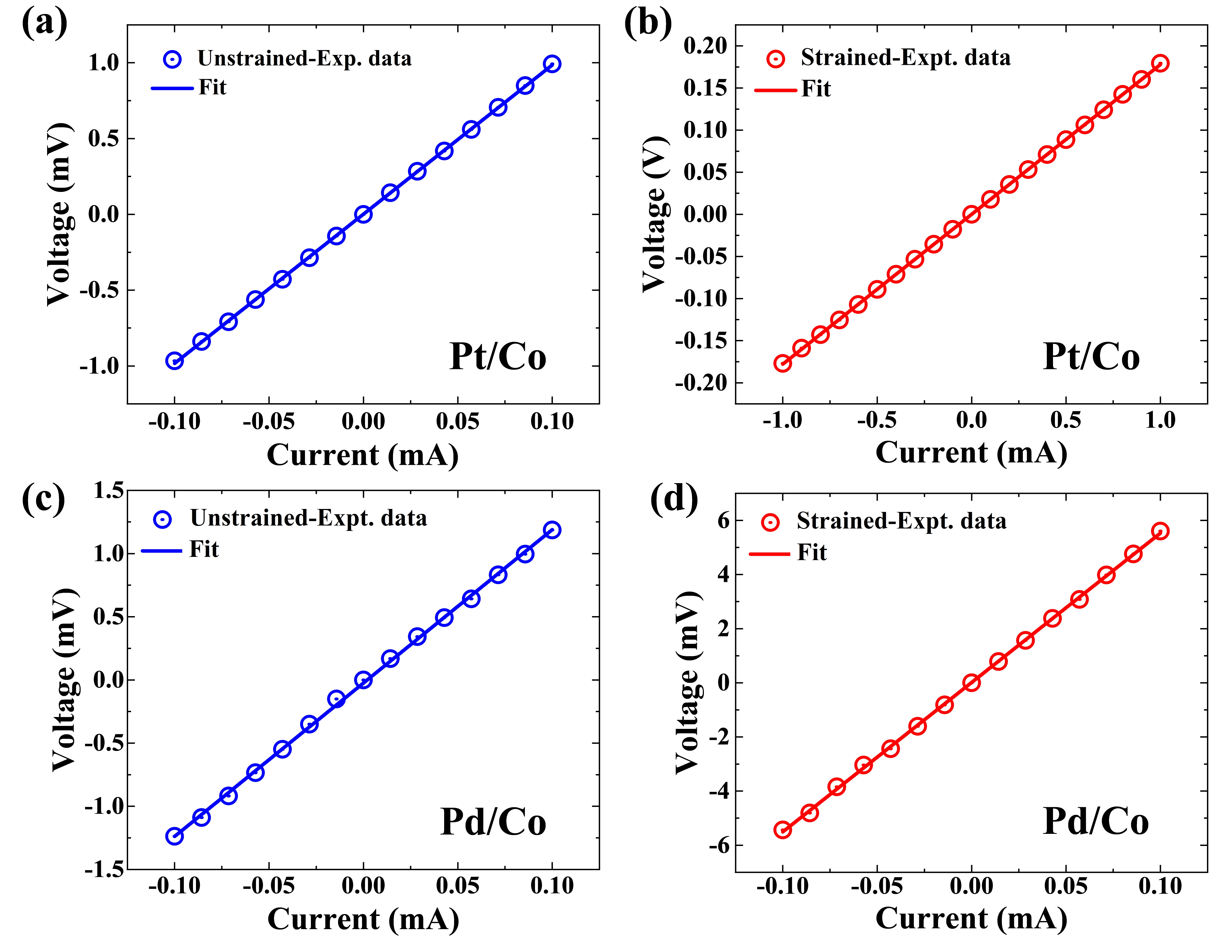}
	\caption{$V$-$I$ curves were measured via four probe techniques for both the Pt/Co and Pd/Co samples, in their (a, c) unstrained and (b, d) strained states. The solid lines are linear fit to the data.}
	\label{fig: Figure_5}
\end{figure}

Hysteresis loops and corresponding domain images are recorded using MOKE-based microscopy in the polar mode. A square-shaped hysteresis loop is observed in the unstrained state of both Pt/Co and Pd/Co samples (see Fig. 6 (a, c)), owing to their moderate PMA strength. The reversal was found to be associated with bubble domain nucleation and subsequent propagation, as depicted in Fig. 6 (b, d). Application of strain increases the coercivity and hence, the switching field of the samples, as shown in Fig. 6 (a, c). This is related to an increased pinning potential due to the damage to the samples. The cracked and buckled features act as additional high pinning barriers and hence, a higher field is required to overcome them. The domain nucleation and propagation are found to be strongly affected by the damage of the film. Due to the periodic buckled features, the bubble domains are not observed in the strained sample. Instead, the domains get elongated and propagate in the transverse direction of the applied stress. Due to the numerous pinning barriers along the stress axis, the domains mainly propagate along the other transverse axis and get an elongated shape. The observed magnetic properties are also found to be irreversible in nature. In an earlier study by Merabtine et al., the authors predicted that the structural discontinuity due to the cracks and buckles should not affect the magnetic properties of a sample \cite{merabtine2018origin}. However, our study contradicts this observation, as the magnetization reversal phenomena get strongly modified by the appearance of such structural discontinuity in the film.  

\begin{figure}[h!]
	\centering
	\includegraphics[width=0.9\linewidth]{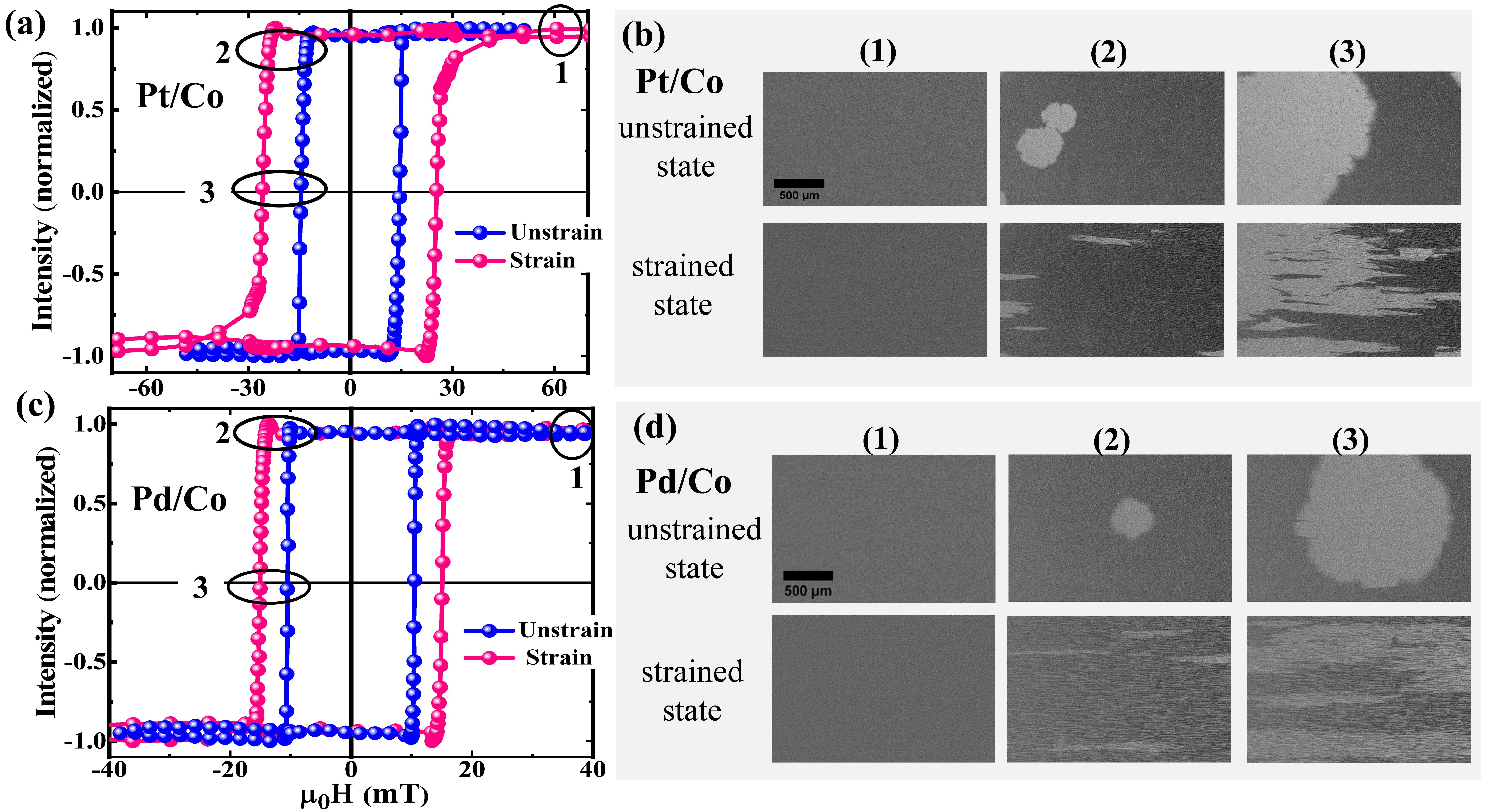}
	\caption{Magnetization reversal is studied in the unstrained and strained states of both the Pt/Co and Pd/Co films via polar MOKE microscopy, where (a, c) hysteresis loops, and (b, d) domain images of both the samples. The images are associated with the points (1, 2, 3) marked in the hysteresis loops. The scale bar shown in the saturation images (1) is 500 $\mu$m in length and is valid for all the other images.}
	\label{fig: Figure_6}
\end{figure}

To better understand the reversal phenomena, we have further studied the magnetization relaxation mechanism at both the strained and unstrained states of Pt/Co and Pd/Co samples. Relaxation measurements are performed to investigate the ability of thermal energy to attain a saturation state under such high strain. In addition, it gives more insight into whether the reversal process is dominated by domain nucleation or DW motion events. Previously magnetization relaxation was studied for nanoparticles, magnetic antidots, and continuous thin films \cite{prozorov1999magnetic,mallick2015size,chowdhury2016study}. Among different proposed models, the Kolmogorov-Avrami model takes care of the inhomogeneities that appear on a sample surface, and thus the relaxation mechanism is explained by a compressed or stretched exponential function \cite{xi2008slow}. The measurement protocol is as follows. First, the sample is saturated by applying a sufficiently high magnetic field and then the field is changed manually to a sub-coercive field value (0.99$H_C$). At that constant applied field (Zeeman energy), the magnetization reversal takes place with the help of the thermal energy. All the domain images are captured during the whole thermal relaxation process, where the amount of dark grey contrast represents a measure of the magnetization in the sample. The intensities of the domain images are extracted using the ImageJ software and the normalized intensity is then plotted with the time taken to complete the reversal. The experimental data is usually fitted using a compressed exponential function written as \cite{xi2008slow},

\begin{equation}
I(t)=I_{1} +I_{2}(1-exp(-(\frac{t}{\tau})^\beta))
\label{2}
\end{equation}

where $I(t)$ is the Kerr intensity measured at time $t$, ($I_1$+$I_2$) is the normalized Kerr intensity, $\beta$ is an exponent that may obtain values between 1 (nucleation-dominated relaxation) to 3 (DW motion-dominated relaxation), and $\tau$ is the relaxation time constant. From the fitting, the dependent parameters $\beta$ and $\tau$ can be extracted, giving more insight into the reversal process. 
\begin{figure}[h!]
	\centering
	\includegraphics[width=0.9\linewidth]{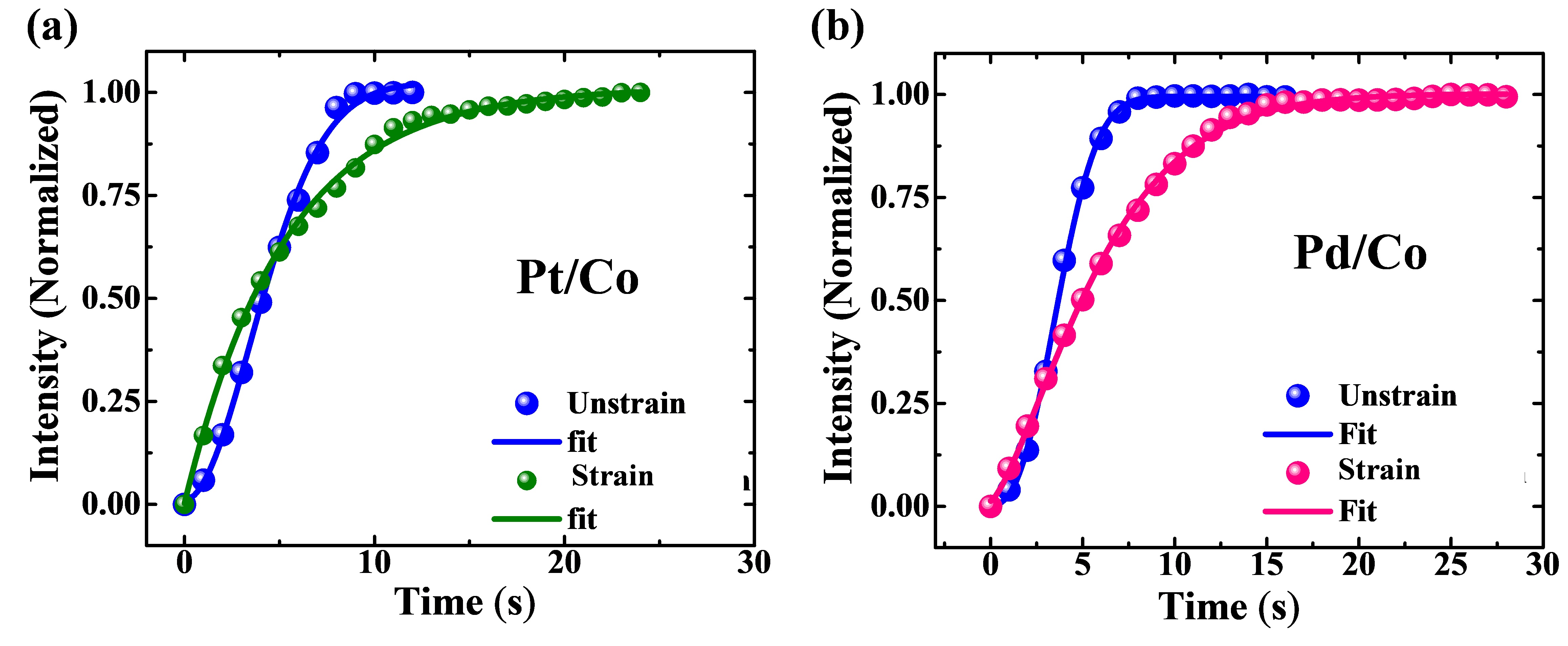}
	\caption{Magnetization relaxation measurement performed using MOKE microscopy in the polar mode. Intensity ($I(t)$) vs. time ($t$) plot is shown at the unstrained and strained states of (a) Pt/Co and (b) Pd/Co samples.}
	\label{fig: Figure_7}
\end{figure}

\begin{figure}[h!]
	\centering
	\includegraphics[width=0.9\linewidth]{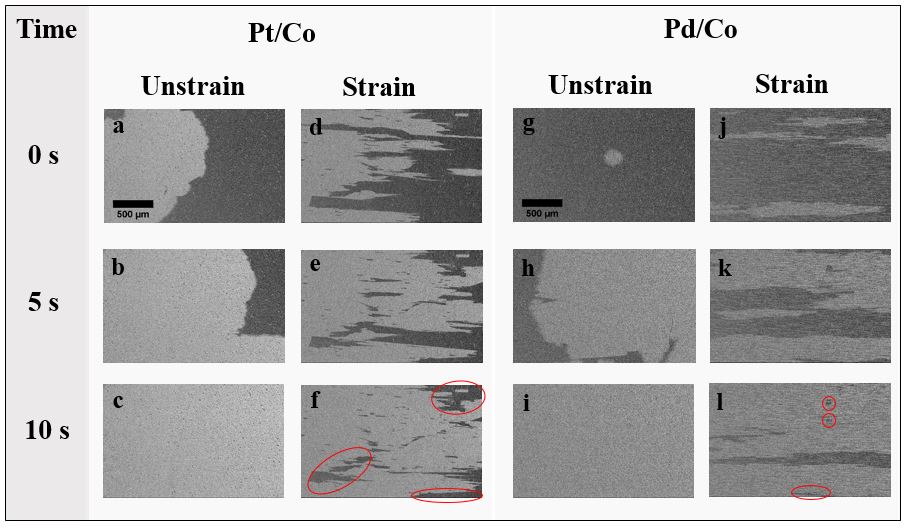}
	\caption{Domain images captured during the relaxation measurement for both Pt/Co and Pd/Co samples. Here, (a-c) unstrained and (d-f) strained states domain images for the Pt/Co film, whereas (g-i) unstrained and (j-l) strained states domain images for the Pd/Co film. The red-marked areas highlight the unreversed regions of the samples. The scale bar shown in the domain image captured at 0s is 500 $\mu$m in length and is valid for all the other images.}
	\label{fig: Figure_8}
\end{figure}

Fig. 7 shows the intensity vs. time plot and Fig. 8 shows the domain images captured at different time intervals during the relaxation measurement. Here 0 s represents the time when the first domain image is captured after keeping the magnetic field at a fixed value. By fitting the experimental data with equation 3 the values of $\beta$ and $\tau$ are extracted at the unstrained states of the samples. The value of $\tau$ is $\sim$ 5s for both the samples in their unstrained states. However, for the strained condition, the saturated state is not reached even after $\sim$ 30s seconds of reversal as shown in Fig. 8. A few unreversed areas were still present due to the structural discontinuities formed in the strained state of the samples. These unreversed areas are marked with red colour in Fig. 8 (f and l). Such high pinning areas seem tough to overcome via the thermal energy and require the application of an additional Zeeman energy in the system. As the sample did not reach magnetic saturation in the strained state, the intensity ($I(t)$) extracted from the domain images does not represent the final saturated state. Thus, the relaxation time constant ($\tau$) extracted from an intensity ($I(t)$) vs. time ($t$) plot, for the strained case, will accompany a finite error, which is not desirable. Thus, the fitted values of $\beta$ and $\tau$ for the strained state should not be compared with the unstrained state. However, by analyzing the domain images and the relaxation curves, it is quite evident that at the strained state, the samples require a higher time to reach saturation in comparison to their unstrained state. Notably, when the sample was in the unstrained state, the reversal was found to be dominated via both domain nucleation and propagation (as $\beta$ obtains a value within 2-2.5). However, after the application of strain, the reversal was mainly dominated by domain nucleation, owing to the several hindrances for DW propagation. Thus, the relaxation measurement was found to be strongly affected by the damage of the film and the thermal energy is not even sufficient to complete the reversal in the presence of a constant Zeeman energy. Thus, the structural discontinuities largely impact the local magnetization reversal, domain, and relaxation dynamics of the flexible film.

\section{Conclusion}

In this work, the impact of thin-film damage on the magnetic properties of flexible Co/Pt and Co/Pd film is investigated. SEM imaging shows that peeling-induced strain generates significant damage (cracking, buckling, etc.) to both films owing to the high elastic mismatch and interfacial defects between the film and substrate. Further, the resistivity measurement shows an increment in sample resistance owing to the peel-off method of the film. Magnetization reversal and domain dynamics are found to be strongly affected by the structural discontinuities originated due to the applied strain. Few pinning barriers are even tough to overcome by the thermal energy as observed during the relaxation measurement. In contrast to bending-induced strain, here the modifications in magnetic properties are found to be irreversible in nature. Thus, the presence of structural discontinuities strongly affects the local magnetic properties (e.g., magnetization reversal, DW dynamics, etc.) whereas the global properties will be affected by the residual stress generated during the damage of the film. Thus our study highlights the impact of the peeling mode of flexibility on both the structural and magnetic properties of flexible PMA films and is expected to attract more research attention on similar aspects.

\begin{acknowledgement}

The authors would like to thank Mr. Brindaban Ojha for his technical help during the manuscript preparation. The authors also thank the Department of Atomic Energy, Govt. of India for providing financial assistance to carry out the research work.
\end{acknowledgement}

\bibliography{references}

\end{document}